\documentclass {elsart}
\usepackage{mathptmx}
\usepackage{amsmath, amscd, amssymb}
\usepackage{mathrsfs}

\journal{Physica A}

\newcommand{\<}{\langle}
\newcommand{\FT}{\Phi^\times}
\newcommand{\F}{{\Phi}}

\newcommand{\ts}{\tilde\psi}

\newcommand{\G}{\Gamma}

\newcommand{\HH}{{\mathscr{H}}}

\newcommand{\rhs}{\F\subset\HH\subset\F^\times}

\newcommand{\R}{{\mathbb R}}
\newcommand{\RR}{{\mathbb R}}
   
\newcommand{\II}{{\mathbb I}}
\newcommand{\C}{{\mathbb C}}

\newcommand{\CS}{{\mathscr{S}}}
\newcommand{\CU}{{\mathscr{U}}}

\newcommand{\g}{{\mathfrak{g}}}

\newcommand{\tauf}{\tau_\F}

\newcommand{\bee}{\begin{enumerate}}
\newcommand{\ene}{\end{enumerate}}
\newcommand{\een}{\end{enumerate}}
\newcommand{\bea}{\begin{eqnarray}}
\newcommand{\eea}{\end{eqnarray}}
\newcommand{\beq}{\protect{\begin{equation}}}
\newcommand{\eeq}{\protect{\end{equation}}}

\begin{document}

\begin{frontmatter}

\title{{Generalized Wavefunctions for  \\
Correlated Quantum Oscillators I: \\
Basic Formalism and Classical Antecedants.}}

\author{S. Maxson}
\address{Department of Physics\\University of Colorado
Denver\\Denver, Colorado 80217\\ 
\ead{steven.maxson@ucdenver.edu}
}

\date{\today}

\begin{abstract}

In this first of a series of four articles, it is shown how a
hamiltonian quantum dynamics can be formulated based on a
generalization of classical probability theory using the
notion of quasi-invariant measures on the classical phase space for 
our description of dynamics.  This is based on certain  
distributions as probability amplitudes, and related distributional 
measures, rather than the familiar invariant Gibbs measures
of classical statistical mechanics.  The 
first quantization is by functorial analytic
continuation of real probability amplitudes, mathematically 
effecting the introduction of correlation between otherwise
independent subsystems, and whose physical consequence is the
incorporation of Breit-Wigner resonances associated to Gamow vectors 
into our description of dynamics.  The real probability amplitudes are 
probably of more formal than practical physical interest, but the demonstrate
structure and the first quantization is functorial. The resulting representation 
of dynamics is indistinguishable from the rigged Hilbert space formulation 
of quantum  mechanics, and this quantum dynamics admits a natural 
field theory interpretation.   This basic formalism will be employed in subsequent
installments of the series.

\end{abstract}

\begin{keyword}{hamiltonian quantum dynamics \sep rigged Hilbert space \sep hamiltonian field theory}

\PACS 11.10Ef \sep 03.65Ca \sep 03.65Db    

\end{keyword}

\end{frontmatter}

\section{Introduction}\label{sec:intro}

\subsection{Motivation}\label{sec:motivation}

This is the first installment of a four installment series of papers
describing various aspects of a hamiltonian description of quantum
fields~\cite{II,III,IV}.  We will create a quantum field theory over
canonical position and momentum (phase space) variables, eventually
adopting the familiar creation and destruction operator formalism when
we deal with the dynamics of correlated (i.e., interacting)
fields.  We will incorporate Schr{\"o}dinger's equation into the
theory, but
extend the allowable solutions to the Schr{\"o}dinger equation to
include weak (distributional) solutions.  This extension of the
allowable solutions necessarily involves mathematical developments
subsequent to von Neumann's Hilbert space (HS) formulation of quantum
mechanics~\cite{vonN}.  Herein, we will avail ourselves of a variant
of the rigged Hilbert space (RHS) formalism of Bohm and
Gadella~\cite{dirackets},
which is a natural generalization of the Hilbert
space formalism of von Neumann that affords us this enlarged
solution set, and which retains the HS formalism of
von Neumann for the description of stationary states.  (See the
reviews~\cite{bgm,bmlg,bh}.)  In this part, then, we will use a
probabilistic description of hamiltonian dynamics on classical phase
space, extended to a quantum theory by the introduction of
multi-tiered notions of correlation.  From the correlation emerges
(Breit-Wigner) resonances, with self-consistent interactions and
non-trivial dynamical time evolution.  Mathematical necessity compels
us to use the RHS formalism, in which our wave functions are ``very
well behaved'', i.e., functions of rapid decrease, generalizing the
notion of Gaussian wave packet.  One of the many interpretations of
this is as a hamiltonian field theory.  

The present series of papers is focused on four aspects of enlarging
the RHS formalism of Bohm and Gadella:
\begin{itemize}
\item{Adding solutions to concrete problems (which cannot even be
respectably addressed in the conventional HS formalism) to the RHS's
formal achievements in the area of intrinsic irreversibility
(including intrinsic irreversibility on the microphysical
level~\cite{qat1,qat2,qat3}) and causality~\cite{bh}.}
\item{Demonstrating how it is possible using our variant of
the RHS quantum formalisms
to incorporate genuine mathematical hyperbolicity into quantum
dynamics.  This leads to notions of quantum dynamics which have many
formal mathematical attributes analogous to the non-linear dynamics,
chaos, fractals, etc., encountered in the contemporary literature for
classical dynamical systems.  This is
categorically incompatible with von Neumann's HS formalism.This will
be covered in installment three~\cite{III}.}
\item{The variant of the RHS formalism which we adopt admits a field
theory interpretation, and from the dynamics alone of 2, 3 and 4
correlated fields respectively we will deduce Yang-Mills structures.
These "generic" gauge field theories  will be covered in installment 
four~\cite{IV}.}
\item{We will require a complex plane structure in our ring of
  scalars, making the algebraic field $\C$ unsuitable for several technical
  mathematical reasons, including uniqueness.  This will be covered in
  length in installment two~\cite{II}. There we will deal with matching
  a substantial number of mathematical structures into a compatible and 
  well defined whole.} 
\end{itemize}
The concrete problem addressed throughout these four papers 
is that of describing the dynamical evolution of
coupled (correlated) quantum oscillators, which is intrinsically
interesting from a physical perspective.  This is a first description of 
this problem from a novel physical and mathematical perspective, and 
there are some quite deep conceptual concerns we will
touch on--we will try to illuminate many of these, but there is no
pretense of complete and definitive resolution on many key points.
Certainly there is much of at least formal interest, and much
structure will be revealed which is physically illuminating even if
one does not take our variant of the RHS picture completely to heart, 
particularly the idiosyncratic real probability amplitude description 
whose analytic continuation introduces a symplectic structure and  thereby
works a quantization.  Whether
or not our developments are physically correct will have to await
later judgment, but our conclusions are largely generic, with a very
strong element of first principle deduction independent of the
physical 
makeup of the fields or the nature of their interaction.  

For instance, we will ultimately conclude that the gauge group of the
Standard Model is very similar (even isomorphic)
to the most general gauge structure one 
might expect for three fields, independent of the precise nature of
those fields and determined solely by dynamics, but that the true gauge
group must be exact.  Thus, our developments
will lead to essentially the same spectrum--isomorphic gauge groups--as 
the Standard Model, but there are differences in interpretation, such as dynamical 
symmetry breaking. If nature does not observe     
the precise ``generic'' gauge structure we predict for a given number of fields, 
then this suggests some special or non-generic physics is present, the possible
existence of a greater or lesser number of fields, etc.

This first installment is concerned with establishing that a well
defined mathematical formalism exists for our constructions, and 
in an aside we will show 
that analytic continuation of a real probability theory  could possibly  
produce a structure which looks just like quantum theory, but 
our resulting theory admits a 
hamiltonian field theoretic interpretation (for correlated fields) whether or 
not one adopts this particular perspective.  

The second
installment will demonstrate the mathematical steps which must be
taken in order to have mathematically well grounded complex spectra:
there are quite well defined points of departure from the usual
Hilbert space quantum formalism and also from the ``standard''
rigged Hilbert space
(distributional) generalization, although this rigged Hilbert space
(RHS) formalism must also be parallelized in many regards in order to
accommodate the complex energies of Breit-Wigner resonances by
mathematically well defined means.  

The third installment illustrates how full fledged hyperbolic dynamics 
may be represented in the quantum mechanics and
hamiltonian quantum field theory so
formulated--the exponential decay of Breit-Wigner resonances is an
example of exponential separation of trajectories, and may be
associated with an increase in entropy.  

The fourth and concluding installment of this series of papers ties
up a lot of mathematical loose ends by defining a Clifford algebra
structure in which our spinorial constructions are mathematically well defined,
and demonstrating how a mathematically well defined gauge theory may
be extracted from pretty much generic considerations of dynamics
alone.  This results in a largely generic construction, with
the gauge group being dependent only on the number of fields, and
recovers an exact gauge group in all cases, and the dynamical
mechanisms for symmetry breaking are shown to exist beyond this most
foundational and elemental gauge symmetry level. In this concluding paper
we will show how the 
islands of relative stability we call elementary particles are
related by an exact symmetry, but their full dynamical interactions are 
described by a larger {{\em {semi}}group which may produce unstable
resonances. It will be shown that the 
gauge structure for four fields over canonical variables may be
associated with structures in a non-trivial space-time with
$(+,-,-,-)$ local signature, in a manner which suggests that some measure of Lorentz 
invariance may be a consequence of dynamics and even suggests
that there is a dynamical basis for PCT.
There is an association of hamiltonian dynamics of canonical
variables, structures on non-trivial space-time, and possibly even to 
irreps of the inhomogeneous Lorentz group, although our group and Lie algebra
representations are fundamental representations (i.e., spinorial) and not
UIR's.  

So this is where a probabilistic description
of canonical dynamics on phase space, in a formalism which 
includes multiple levels
of correlation, will ultimately lead us.  At a minimum, we will arrive
at a fiber bundle description of {\em all} canonical dynamics
possessing analytical interactions, which lead to stable correlated
structures, and which can be phrased in a probability formalism, i.e.,
based on ensemble notions.  This bundle structure will relate
the various stable structures present in the dynamical system, while
there is a larger dynamical structure associated with the production
and subsequent dynamical evolution of resonances not amenable to the
fiber bundle formalism, being based on semigroups rather
than full group structures.  See installment four~\cite{IV}.  
Perhaps there is some particle physics in here, but, for instance,
the basic results should be just as applicable to fluids or plasmas
with long range interactions.

\subsection{The present paper}\label{sec:present}

The canonical transformations of physics are the symplectic
transformations of mathematics.  The second paper of this series will choose to 
represent the appropriate symplectic (semi-)group as the largest possible
dynamical (semi-)group, generalizing those classical notions.
In the theory of dynamical systems,
these symplectic transformations are the dynamical transformations.
See~\cite{souriau}.  Classically, hamiltonian dynamical evolution is
associated with the 
translation and reshaping of a region in phase space to which a
dynamical system is localized, and this dynamical evolution is by 
symplectic (e.g., ``area preserving'') maps. These are topologically transitive
mappings. 

We will use a 
sophisticated mathematical structure herein, whose
implications we begin to relate in Section~\ref{sec:structure}.  For
instance, we can replace the classical description of a dynamical
system as a point in phase space with a more loosely localized
region in phase space to which
our dynamical system is localized by a ``bump function'' (function of
rapid decrease, e.g., a generalized gaussian) which is
properly normalized so that it may be regarded as a probability
amplitude.  We can then provide a similar ``bump function'' to represent
sensitivities and capture probabilities, etc., of a measurement
apparatus as another probability amplitude. (We might also use a
more exotic generalized function, or distribution, rather than 
a simple function here.)  There will be
symplectic (=dynamical) transformations on the function space made up
of all the mathematically acceptable ``bump functions,'' provided we
give this function space a symplectic structure during its construction. 
(The detailed construction of this symplectic structure is the subject
of our second paper.)  
Those symplectic transformations could possibly
represent dynamical evolution of these classical (i.e., {\em real})
probability amplitudes within the context of a classical probability
theory in which the outcome of a measurement is predicted by the
scalar product of probability amplitudes, and this scalar product
should be able to 
evolve with time: we are able to represent the time dependent
interaction of a non-stationary dynamical system with a non-stationary
measurement apparatus using this formalism.  We will defer
explicit  consideration of the
symplectic transformations until the second installment, and
concentrate on how the nature of the spaces involved lends itself to a
probability theory in this initial installment.  For us, probabilities
will be an intrinsic part of the construction, and not a subsequent
interpretation.

The analytic continuation--a complex symplectic transformation--of just
such a probabilistic description starting with {\em real} probability
amplitudes is mathematically equivalent to representation of quantum
resonances, including Breit-Wigner resonance poles (of the analytically
continued S-matrix) belonging to one of the spaces of a Gel'fand
triplet in our RHS formalism. Additionally, this accords--perhaps only
loosely at times--with the physical 
interpretation of the RHS formalism in~\cite{bgm}.  See
Section~\ref{sec:classical}. 

Our quantum theory will be obtained then by a functorial analytic
continuation of classical densities on phase space, and our
measurement apparatus throughout will be associated with distributions 
mathematically dual to those densities.

\section{Structure of the Spaces Involved}\label{sec:structure}

Working physicists virtually never use the full generality of the
Hilbert space of von Nuemann's formulation of quantum mechanics.
Typically, they employ smooth functions and Riemann integrals rather
than classes of Lebesque square integrable functions.  Although the
smooth functions used by physicists usually do belong to the Hilbert
space, they are almost 
invariably elements of the Schwartz space $\CS$ of functions of rapid
decrease, so that, e.g., there are vanishingly small contributions to
the probability attributable to infinities in energy or momentum or
position.  The rigged Hilbert space (RHS) formalism, which we shall
use a variant of, adopts this usual practice of working physicists,
and obtains thereby a variety of additional benefits, such as an
enlarged class of solutions to the Schr{\"o}dinger equation.  We will
be exploring the consequences of adopting an enlarged class of
solutions to the Schr{\"o}dinger equation in a RHS.  (The ``rigging''
in rigged Hilbert space is based on a nautical methphor: one thereby
makes the bare Hilbert space ``ready to sail''.)  There are several
excellent reviews on the RHS formalism which describe the Gadella
diagrams we will make use of below~\cite{bgm,bmlg}, and one which
addresses many issues including causality and the mathematics
summarized in the Gadella diagrams (although not using the diagrams
themselves)~\cite{bh}. 

There has been a recent revival of interest in 
working in a quantum paradigm involving use of generalized
wave functions in a rigged Hilbert space (RHS, which will be indicated
generically by the Gel'fand triplet of spaces $\rhs$).  This RHS
formalism involves applying
a mathematically rigorous methodology of analytic continuation, 
and leads to the unification of the vector space representation of 
quantum systems and the representation of resonances by poles of the 
analytically continued S-matrix.    In the RHS paradigm, 
resonances are represented by abstract Gamow vectors associated to
Breit-Wigner  
poles of the analytically continued S-matrix, with exponential decay, 
and the time evolution is by semi-groups.  These semi-groups provide the 
formalism with the ability to express the 
boundary conditions of an irreversible physical process without
regeneration \cite{qat1,qat2,qat3}.  

We follow the lead of Gadella~\cite{gadella}, who
refined the analysis issues in the function spaces relevant to our
work, 
and showed how to obtain a rigorous analytic continuation 
based on necessary and sufficient conditions.  Our concern, then, is
erecting particular geometric structures in the types
of spaces Gadella has shown us we must use.
We begin by very tersely describing the
interrelationships of the spaces involved in the RHS formalism, and
also providing references for further inquiry.  The RHS formalism
involves working with a hierarchy of triplets of spaces, with abstract
spaces and function space realizations of the abstract
spaces.

Distinguishing abstract spaces from their function space
realizations, and the analytic continuation of those spaces, 
is an essential, if tedious part of the RHS methodology.
This is physically
motivated in part.  There is considerable structure in the large
number of vector spaces involved, and there is a significantly
different physical content to each of the spaces: there is a
difference between the abstract Hilbert space ($\HH$) and its $L^2$
function space realization in the energy representation on the
half-line, $L^2[0,\infty
)$, and the space corresponding to the abstract $\HH$ realized
in a subspace of $L^2 (\R )$, also in the energy 
representation.  Whether or not the energy spectrum is bounded from
below has considerable practical physical implication 
as well as formal mathematical importance.

Formally, the RHS paradigm involves alternative topological
completions of a pre-Hilbert space, the algebraic linear space $\Psi$,
to form the Gel'fand triplet of spaces 
$\rhs$.  In the $\tau_\Phi$ topology set by a countable family of
semi-norms we have one topological completion to form the topological
vector space $\Phi$.  Using
the scalar product to define a norm, we define a Hilbert space
topology, $\tau_\HH$, and the completion of $\Psi$ in this topology is
a Hilbert space $\HH$~\cite{note0x5}.  Dual to $\Phi$ is the conjugate
space $\Phi^\times$, which has a weak-dual topology, $\tau^\times$.
The resonances in the RHS formalism are obtained by analytic
continuation, which proceeds from
$\HH^\times\longrightarrow\tilde\Phi^\times$ \cite{gadella}, where
$\HH^\times\cong\HH$ (Riesz isomorphism), and $\tilde\Phi^\times$ is
the complex extension (analytic continuation) of $\Phi^\times$.  There
are complex energy eigenvalues 
on $\tilde\Phi^\times$, indicating exponential decay for quantum
systems represented by states there.  In terms of (mathematically
respectable) physical content, in the RHS formalism we have the
analytically continued S-matrix (and its poles), the Lippman-Schwinger 
equation, M{\o}ller operators, etc.  Although we will not deal with
these explicitly, in installment two~\cite{II} we will effectively see 
M{\/o}ller wave operators as dynamical (=symplectic) transformations,
an algebraic Lippman-Schwinger equation, etc.  As a point of
distinction from prior work, previously the analytic continuation
(complex extension) was of the absolutely continuous scattering
spectrum, while the present work operates by complex extension
(complex symplectic transformation) of an operator algebra in which
the only spectrum initially apparent is discrete.  See installment
two~\cite{II}.  We will not deal closely with representation issues, but
will implicitly be working with both the operator algebra and its
representations on both abstract and function spaces throughout.
Even where we are not explicit in this series of articles, it is always
implied that we intend eventually to work with representations, make
use of spectral theorems, and so on.

Although we will not speak explicitly of
dragging contours around in the complex plane, or making explicit use
of the residue theorem, etc., nothing we will do will change the
necessary and sufficient conditions for the complex
extension~\cite{gadella}, or any of the other major structural
features of the RHS formalism.  (See the reviews~\cite{bmlg,bgm} for
more details of that formalism than it is possible to recapitulate
here.)  In installment 
two~\cite{II}, by rotations (and possibly other complex symplectic
transformations), we will shift the spectrum (i.e., the poles of the 
resolvent) from the real axis out into the complex plane.  Only
a certain countable set of complex symplectic transformations will
produce the Breit-Wigner resonance poles.

The basic structure of the relationships of the abstract spaces and the
associated function spaces which occur in the RHS formalism are most
easily seen in the Gadella diagrams~\cite{bgm,bmlg}.  Conventionally one
envisions a measurement process having the
preparation procedure ending at time $t=0$, with preparation of the
in-state $\phi^+$ occurring during $t\le 0$ and observation of the effect
$\psi^-$ during $t\ge 0$.  The Gadella diagram for representing
the preparation of an in-state $\phi^+$ during $t\le 0$ is
\begin{equation}
\begin{CD}
{\phi^+\in}\;\; @.\F_-\subset@.\HH\subset@.(\F_-)^\times@.
{\ni{\ts^G} =|E-z^\ast_R \;^+ >}\\
@.        @V\CU^- VV  @VV\CU^- V   @VV(\CU^-)^\times V@.      \\
{\<^+E|\phi^+\rangle\in}\qquad@.\CS\cap\HH^2_- \Bigm|_{\R^+}\subset@.
L^2[0,\infty)\subset@.
\Bigl(\CS\cap\HH^2_- \Bigm|_{\R^+}\Bigr)^\times\qquad@.      \\
@.        @V(\Theta_-)^{-1} VV  @.    @VV(\Theta_-^\times)^{-1} V@.   \\
        @.\CS\cap\HH^2_-\subset@.\HH^2_-\subset@.(\CS\cap\HH^2_-)^\times@.
{\ni -i\sqrt{\frac{\G}{2\pi }} {\frac{1} {E-z^\ast_R}} }
\label{eq:gadella-in}
\end{CD}
\nonumber
\end{equation}
and the corresponding diagram for the effect $\psi^-$ 
observed during $t\ge 0$ is
\begin{equation}
\begin{CD}
{\psi^-\in} \;\; @.\F_+\subset@.\HH\subset@.(\F_+)^\times@.
{\quad}{\ni \psi^G = |E-z_R \;^-> }\\
@.          @V\CU^+ VV   @VV\CU^+ V     @VV(\CU^+)^\times V@.\\
{\<^-E|\psi^-\rangle\in}\qquad@.\CS\cap\HH^2_+ \Bigm|_{\R^+}\subset@.
L^2[0,\infty)\subset@.\Bigl(\CS\cap\HH^2_+\Bigm|_{\R^+}\Bigr)^\times@.\\
@.      @V(\Theta_+)^{-1} VV @.       @VV(\Theta_+^\times)^{-1} V@.\\
       @.\CS\cap\HH^2_+\subset@.\HH^2_+\subset@.(\CS\cap\HH^2_+\,)^\times@.
{\ni i\sqrt{\frac{\G}{2\pi }} {\frac{1}{E-z_R}}}
\label{eq:gadella-out}
\end{CD}
\nonumber
\end{equation}
The basic ideas behind these stem from~\cite{gadella}; it or~\cite{dirackets} should be
consulted for details of construction for the various spaces and 
mappings, as well as the careful definition and meaning of the various vectors
identified in the left and right hand margins.  Each
level of each diagram contains a RHS or Gel'fand triplet of spaces,
and there are certain properties which each space inherits by virtue
of its relative position in a Gel'fand triplet.

Along the top line of each diagram are abstract spaces, the middle
line gives the function space realizations of the spaces in the top
line conforming to the necessary and sufficient mathematical conditions
for performing analytic continuation in a
unique fashion (energy picture, physical energy bounded from below
with boundary conventionally taken as $E=0$), and the bottom line
shows the function spaces resulting from the analytic
continuation~\cite{gadella}. 
The lack of connecting link between the $L^2 (\R_+)$ realization of
$\HH$ on the middle level and the subspace of
$L^2 (\R )$ on the lower level reflects the
lack of unique extension between these spaces.

The transformation
\begin{equation}
\CU^+ \, : \, \F_+ \ni \psi^- \longrightarrow \langle^- E \vert
\psi^-\rangle \in \CS\cap\HH^2_+\vert_{\RR^+} \;\; .
\label{eq:inmap}
\end{equation}
defines the space $\F_+$ in terms of the function space
$\CS\cap\HH^2_+\vert_{\RR^+}$ (the well-behaved functions on the
positive real line $\RR^+$ which are also Hardy class functions from
above) extends to a unitary transformation $\CU^+ : \HH
\longrightarrow L^2 (\RR^+ )$, i.e., the Hilbert space of Lebesque
square integrable functions on the positive real line
\begin{equation}
(\CU^+ f , \CU^+ g) \equiv \int_0^\infty \; dE \langle f | E^-\rangle
  \langle^- E|g\rangle = \langle f|g\rangle \; , \quad f,g \in \HH
  \,\, .
\label{eq:inunitary}
\end{equation}
The map $(\CU^+ )^\times$ is the extension of $(\CU^+ )^\dag$.  Similar
relations hold for the transformation
\begin{equation}
\CU^- \, : \, \F_- \ni \phi^+ \longrightarrow \langle^+ E \vert
\phi^+ \rangle \in \CS\cap\HH^2_-\vert_{\RR^+} \;\; .
\label{eq:outmap}
\end{equation}
The hardy class functions have the remarkable property that their
values on the entire real line $\RR$ are determined by their values on
$\RR^+$.  In the second installment~\cite{II}, we will consider the
symplectic 
transformations (=dynamical transformations), which form a superset of
the unitary transformations, and their action on the spaces of
states will be defined so that they obey the same rule as that laid
out in equation (\ref{eq:inunitary}), since that is the form which is
necessary and sufficient for their action on the space to be
symplectic (=dynamical)~\cite{porteous}.  The significance of this is
that one may define the maps in such a way that ones dynamics induce 
unitary 
transformations on Hilbert space, which we will see relates to the
issue of whether or not the transformations are ergodic (installment
three~\cite{III}.)  We will note the diagrams are commutative, and refer
further inquiries by the reader to the references cited.  

Considering only the top diagram briefly for some further
interpretation, we have an abstract
(exponential formation culminating at $t=0$) Gamow
vector $\ts^G$ belonging to the rightmost abstract space $(\F
)^\times$ on the top line being associated to a Breit-Wigner 
resonance pole in the the rightmost
function spaces of the lowest line, which has a complex energy
eigenvalue $z^\ast_R =
E_R + i\Gamma /2$.  As a practical matter, the physically 
prepared abstract state denoted $\phi^+$
is input into this hierarchy as a ``very well behaved'' element of the
function space lying in the intersection of the Schwartz and Hardy
class functions (from below) over the half line 
$\CS\cap\HH^2_- \Bigm|_{\R^+}$, e.g., as a physically determined
``very well behaved'' energy
distribution of a beam actually prepared in an accelerator $\langle^+ E| 
\phi^+ \rangle = \phi^+ (E)$.  The $\R_+$ indicated with the spaces
corresponds to the physical energy
spectrum, which is bounded from below.  The abstract
prepared state $\phi^+$ ($+$ superscript physical notation) is an
represented by an element of the
space of Hardy class functions over the half line from below, $\HH^2_-
\Bigm|_{\R_+}$ ($-$ subscript mathematical notation).  

The second diagram for $t\ge o$ is similar, with an abstract Gamow
vector of pure exponential decay associated with a Breit-Wigner
resonance pole whose complex energy is  $z_R =
E_R - i\Gamma /2$, etc.

In the sequel, whether the given use of 
a rigged Hilbert space $\rhs$ is intended generically or as a
particular Gel'fand triplet of spaces often
will depend on the context of use.  The spaces usually will be indicated as 
$\F$ or as $\F_{\g\pm}$, or as $\subset \CS\cap\HH^2_\pm$, etc., 
depending on whether one is concerned with a generic rigged Hilbert space
structure (abstract or function space realized), 
its Lie algebra representation structure, or the space of 
``very well behaved'' vectors in the 
intersection of the Schwartz and Hardy class functions (from above or
below).

\section{A classical system extended}\label{sec:classical}

For a real elliptic Hamiltonian (which the analytic continuation 
procedure must start with, see installment three~\cite{III})--such as
the Hamiltonian of a system of  two free oscillators--there exist {\em real}
eigenfunctions.  Indeed, this is the way the function space
realizations of the vectors of quantum physics are usually encountered
in the classroom for the first time, and it is only later  
that issues of time evolution and phase freedom for these special 
functions are discussed in detail.  From a
mathematical perspective, it would be reasonable to consider the
abstract spaces $\F_\pm$as real, and to also consider only the purely
real subspaces of the
associated (complex) function spaces identified in the preceding Gadella
diagrams.  See also~\cite{note10}.  

In order to understand some aspects of the present treatment of this
analytic continuation of a Lie algebra representation, it is useful to
adopt an idiosynchratic perspective, and to consider extending the
real representation of the real algebra to a representation of its
complexification, both in terms of the abstract representation space
of a group whose Lie algebra is $\g$, $\F_{\g\pm}$, and the proper
function  space realization lying in the
intersection of the Schwartz and Hardy class functions.  At first, this
must be 
regarded as idiosynchratic since the exponential map need not be a
unitary transformation on a real Hilbert space if one follows
the von Neumann model of construction, so that 
probabilities appear not to be Noetherian conserved quantities of the 
evolutionary flow there, etc.  This associated 
{\em real} Hilbert space may not seem very interesting
physically at first, but will offer more promise once it is clear we use
a different Hilbert space than the one von Neumann constructs . Further,  
nonconservation of the probability of observing a
resonance--essentially, its survival probability--suggests decay of the 
non-surviving system, which is not repugnant to us in any way.  We
require the conservation of the {\em total} probability, also
including the 
probability of the system which comes into being as the product of the 
decay of the resonance system which does not survive.  (We may not be 
able to show total probability conservation constructively, but we
will be able to 
show the existence of the equilibrium towards which the system decays
in the third installment of this series, from which conservation of
total probability may be inferred.)

This idiosynchratic perspective arises simply from noting that there
is nothing in Gadella's construction which requires the top two levels
of the Gadella diagrams to involve complex spaces!   There is a field
of classical mechanics which can be called distribution density
dynamics.  (See, e.g., the appendix 14 to~\cite{arnold} and references
therein.)  There is thus a physical context in which the
mathematically permissable choice of using real spaces on the top two
lines of the Gadella diagrams makes sense.  
We can define a classical dynamical system on real spaces $\F_\pm$ and
$\F^\times_\pm$, and the associated physical context is, in effect, 
an averaging over an infinite number of classical trajectories 
(e.g., in phase space) using a distribution density.  Such a 
construction would have abstract spaces and
function space realizations, corresponding to the top two levels of
our idiosynchratic Gadella diagram.  The transition from the middle to
the bottom levels of the idiosynchratic diagram
via analytic continuation represents the (functorial)
first quantization of the classical distribution density dynamics
method of description of a dynamical system, and results in 
the description of a dynamical system by 
a recogniable quantum theory which includes Breit-Wigner
resonances~\cite{note11}.  

Most recent physics using the RHS formalism has been preoccupied with the
energy representation, since the primary recent interest has been in
describing irreversible time evolution using the formalism's ability
to express the boundary conditions of an irreversible process, without
regeneration~\cite{qat1,qat2,qat3}.  There is also a momentum space
picture of the Gamow 
states~\cite{handm,mandh}.  For our present idiosynchratic purposes, it
is interesting to consider starting from a classical phase space, since
when we do so many similarities appear between the RHS methodology
advanced herein and a version of classical statistical mechanics.  

By working with probability amplitudes in the RHS
format, rather than with probability amplitudes on the constant energy
surfaces in phase space, we are defining probability measures on phase 
space itself.  When we cause those probability amplitudes to evolve
dynamically, we avoid the many of the
pitfalls of time evolving probabilities in
the Boltzmann and Gibbs approaches based on discrete
partitioning of phase space.  (See, e.g.,~\cite{italian}.)  

In place of the energy representation of a prepared state $\langle E 
|\phi^{in}\rangle = \langle^+ E |\phi^+\rangle \in\CS \cap\HH^2_-
\Bigm|_{\R^+}$, we will introduce a similarly well behaved $\langle \pi |
\phi^{in}\rangle = \langle^+\pi | \phi^+\rangle \in \overline{
\CS\cap\HH^2_-}$, where $\pi$ indicates the canonical phase space
variables $(p_x, p_y, q_x, q_y)$, and $\overline{\CS\cap\HH^2_-}$
indicates real valued functions on $(T^\times\R^2 ) \oplus i \circ
(T^\times\R^2 )$ (see~\cite{II}),
which are understood to be properly normalized as probability
amplitudes.  For example, an equilibrium ideal gas in a
container, $\langle^+\pi | \phi^+\rangle$ might have 
Maxwellian (gaussian) velocity distributions and a functions of compact
support over the position as the components of a multi-component spinor
(see~\cite{II}).   

Similarly, in place of $\langle E|\psi^{out}\rangle = \langle^-
E|\psi^-\rangle \in \CS\cap \HH^2_+ \Bigm|_{\R^+}$ for the measurement
resolution of our measuring apparatus (i.e., a measure on phase
space), we will have $\langle \pi |\psi^{out}\rangle = \langle^-
\pi  |
\psi^-\rangle \in\overline{\CS\cap\HH^2_+}$, meaning a real valued and
very well behaved function of our $\pi$-variables, also
normalized.
Both of these sets of generalized functions will be defined for
restricted time 
domains just as the functions on the energy surfaces within these
function spaces have restricted time domain of definition, and
dynamical evolution (including dynamical time evolution) will be by
semigroups, meaning we continue to have the ability to express the
boundary conditions of an irreversible process referred to earlier.
(Because we include distributions, we have the conceptual ability to
include ``distributional probability measures'', such as measures
which will yield Boolean value 0 or 1 representing the outcome of a
yes-no experiment.)  Because of the hyperbolic structure inherent on
properly constructed
complex spaces, we we later see it is possible to construct hyperbolic
probability measures, e.g., the probability of observing a
Breit-Wigner resonance decreases hyperbolically with time.

This formalism is also compatible with the notion of extended objects,
because, e.g., $\delta x \ne 0$ in general, so no point
localization is assumed anywhere, although we do have the mathematical
machinery (Dirac measures) to accomodate point localization if
desired.  The idiosynchratic perspective contemplates concurrent 
position and velocity measurements, and this is perfectly okay--we're
not in Hilbert space anymore!  (This is made formally apparent in~\cite{dirackets}.) 
If we had a gaussian classical momentum distribution (e.g., a 
a Maxwellian velocity distribution)
and gaussian classical position distributions, then we would
recover a classical equivalent of the Heisenberg uncertainty relation,
based on the widths of the gaussians,
even in the case that Planck's constant and the full implications 
of the quantum mechanical
complementarity principle are not incorporated into the system.

From this $\pi$-representation in terms of $(p,q)$, analytic
continuation takes us $(p,q) \longmapsto (p,ip,q,iq)$. Following a
simple change of coordinates
\begin{equation}
A = ( q+ip)/\sqrt{2} \qquad A^\dag = (q-ip)/\sqrt{2}
\label{eq:cdops}
\end{equation}
so that now $(p,ip,q,iq) \longmapsto (A,iA,A^\dag ,iA^\dag )$ provides
a basis for the complexified phase space.  We are thus an eyeblink
away from the creation and destruction operator formalism used in the
rest of this series of papers.  In installment four, we will see how
obtaining a  real Witt basis from this basis
enables the representation of either bosons or
fermions without a change of basis (spinors are notoriously basis
dependent).

Joint position and momentum probability distributions do not exist for
any quantum state represented by an element of $L^2 (\R^n )$. (This is the
motivation for the Wigner transform, for instance.)  Both position and
momentum cannot simultaneously be represented by continuous operators
on $L^2 (\R^n )$.  Similarly, on $L^2 (\R^n )$ it cannot be the case
that both the creation and destruction operators are both continuous
operators, and so, for instance, 
one has to use care in defining the coherent states.
However, in the RHS formalism position and momentum are both
represented by operators which are $\tau_\Phi$-continuous and
$\tau_\Phi$-closed.  Likewise, both creation and
destruction operator are $\tau_\Phi$-continuous and $\tau_\Phi$-closed
operators (see chapter two of~\cite{dirackets}), and so 
these restrictions applicable to the HS formalism do not constrain us
in a RHS formalism.  Thus, after the analytic
continuation we change basis to the real Witt basis and think in
terms of eigenvector probability densities of the creation and
destruction operators, since they are $\tau_\Phi$-continuous
$\tau_\Phi$-closed operators.  Continuity and closedness in $\Phi$ and
$\HH$ need not agree, and this has far reaching consequences, which we
are exploring.

\section{Quasi-Invariant Measures}\label{sec:quasiinvariant}

The space $\FT$ provides quasi-invariant measures for the space $\F$ in
the Gel'fand triplet $\rhs$~\cite{genfun4}.  Because we use operators
which are not symmetric, the left and right quasi-invariant measures
will differ, e.g., in the resolution of the identity provided by the
spectral theorem, the dyads $\vert \phi\rangle \langle {{\phi}}
\vert$ are not symmetric, and in particular $\vert \phi\rangle$ and 
$\langle {{\phi}}\vert$ will be defined for different time
domains as a consequence of our use of semigroups of dynamical time
evolution.  This will have physical consequences we will demonstrate
in installment three~\cite{III}.  In installment
three~\cite{III}, we will see that dynamical evolution can be
hyperbolic, and these quasi-invariant measures are hyperbolicly
evolving as well, but that overall probability is conserved.

\section{Physical Interpretation}\label{sec:interpretation}

In the present series of articles, we are exploring a dynamical system
of oscillators in which there is correlation between the oscillators.
We are using probability amplitudes rather than point particle
localization, and in particular we allow the use of densities and
distributions for those probability amplitudes.  This involves us with
not the $L^2$ functions but with a subset of 
the Hardy class functions which
partition $L^2$, $L^2 = \HH^2_+ \oplus \HH^2_-$, according to the
Paley-Wiener theorem.  

Given the possibility of free oscillators, we expect some sort of
direct sum structure (sort of like a Foch space), except our dynamical
transformations may also mix the components.  We will see in
installment two~\cite{II} that these multi-component probability
amplitudes are of mathematical necessity of a certain form and a
sufficient form for them is put forward.  On our phase space, and on
the abstract and function space representations of it (using these
densities and distributions), there is a symmetric form $Q$ and a
skew-symmetric form $J$:
\begin{equation}
Q= \begin{pmatrix} 0 & \II \\ \II & 0 \end{pmatrix} \qquad \textrm{and} \qquad 
J= \begin{pmatrix} 0 & \II \\  -\II & 0  \end{pmatrix}  \quad .
\label{eq:qjmatrices}
\end{equation}
Each of these forms induces a scalar product, so that there are, in
effect, three possible scalar products.  Assuming 
\begin{equation}
\vert \phi ) = \begin{pmatrix} \phi_1 \\ \phi_2 \end{pmatrix} \quad
\textrm{and} \quad 
\vert \psi ) = \begin{pmatrix} \psi_1 \\ \psi_2 \end{pmatrix} 
\label{eq:vectors}
\end{equation}
we may represent these three scalar products as:
\bea
(\psi | \phi ) &=& \psi_1^\times \circ\phi_1 + \psi_2^\times \circ\phi_2 \nonumber \\
\langle \psi | \phi \rangle &=& ( Q \psi | \phi ) = \psi_2 ^\times \circ
     \phi_1 + \psi_1^\times \circ\phi_2  \nonumber   \\
\{ \psi | \phi \} &=& ( J \psi | \phi ) = \psi_2^\times \circ\phi_1 -
     \psi_1^\times \circ\phi_2  
\label{eq:3scalar}
\eea
where $\psi_2^\times  \circ \phi_1$ indicates a ``single component'' scalar
product.  We will return to these alternative scalar products (and their associations
with alternative topologies) in installment four~\cite{IV}, but
the important ovservation for now is the types of correlation shown in
the above alternative scalar products.  Thus, the complex unit scalar
$i$ is associated with a symplectic form $J$, and is thus identified
with a correlation mapping (a type of injective embedding in the
dual)~\cite{porteous}, and in fact a complex plane structure may be
associated with both the symmetric and skew-symmetric correlations
above, depending on whether one adopts an orthogonal or hyperbolic
(Lobachevsky) geometry.

\subsection{Collateral Implications}\label{sec:collateral}

The foregoing ``idiosynchratic perspective'' suggests at once fairly
straightforward and conceptually obvious modifications are likely possible to
the Grad moment expansion~\cite{grad,muller}, but such pursuits are very
wide of our present concerns--possibly it will be taken up another day,
establishing a connection between the present formalism and a
generalized ``thermodynamics''.  In place of a moment expansion about
the Maxwellian (e.g., gaussian) velocity distribution, we generalize
to functions of rapid decrease, and similarly interpret the moments as
thermodynamic quantities.  This also has implications for the Meyer
cluster expansion and its graphical representation by Feynman
diagrams.

There are also implictions for the theory of solution of differential
equations.  Thus, the second installment is concerned with the
Lie-Poisson bracket of vector fields on phase space, and representations
of this structure, and the fourth
installment is concerned with the tensor algebra of phase space.

There is a breathtakingly direct analogy between a
probabilistically oriented classical description and the mathematics
of the present formalism.  We find a description of the classical
treatment in~\cite{arnold} Appendix 14, page 457:
\begin{quote}
``Jacobi realized that the (classical) Poisson brackets of the first
integrals of any hamiltonian system could be considered as a Poisson
structure [reference in original].

The construction of a Poisson structure on the dual space of a Lie
algebra leads to a new Lie algebra.  This construction may then be
repeated, leading to a whole series  of new (infinite dimensional)
Poisson structures.  More generally, suppose that one is given any
Poisson structure on a manifold.  Then the space of functions on that
manifold carries the structure of a Lie algebra.  This implies that
the dual space of this function space carries its own Poisson
structure.  Elements of this dual space may be interpreted as
distribution densities on the original manifold.  Thus, the space of
distributions on a Poisson manifold (for example, on a symplectic
phase space) has a natural Poisson structure.  This structure makes it
possible to apply the hamiltonian formalism to equations of Vlasov
type, which describe the evolution of distributions of particles in
phase space under the action of a field which is consistent with the
particles themselves.''
\end{quote}

Physically, this means we may be working in a paradigm whose
classical (pre-analytic continuation) analogue is based on ensemble
notions, e.g., one is either dealing with an ensemble of particles
(perhaps even a field) or with an ensemble of measurements of
identical simple systems.  In particular, both before and after
anayltic continuation this construction includes 
a field theoretic treatment of a large number of
particles whose paths are locally hamiltonian, and whose interactions
are dealt with in a self consistent way~\cite{note3}. There are both
orthogonal and symplectic correlations, and, in particular, there are both
real symplectic and complex symplectic correlations, corresponding to
corrrelated classical dynamics and to a complex dynamical correlation
which admits oscillatory behavior in a 
formalism which looks just like a quantum theory of resonances.

\appendix

\section{Complex Spectral Theorem}\label{sec:compspectthm}

From the starting point of a real semisimple Lie algebra, we
undertake to show the role of the complex covering algebra in the 
construction of a representation space $\F$.  From the universal embedding 
algebra, one has the 
existence of a complete set of commuting operators.  Let us assume this 
c.s.c.o. of essentially self adjoint operators is
$\{ A_1, \, A_2,\, \dots , A_N\}$.  Let $\Lambda_i$ be the Hilbert space
spectrum of the operator $A_i$, $i =1,\, 2, \dots ,N$, and let
$\Lambda = \Lambda_1 \times \Lambda_2\times \cdots \times \Lambda_N$
be the Cartesian product of the $\Lambda_i$.  Then the general 
Gel'fand-Maurin Theorem (also called general Nuclear Spectral Theorem) 
asserts there exists a rigged Hilbert space $\rhs$ such that there exists
a uniquely defined positive measure on $\Lambda$ such that~\cite{bgm}:
\begin{enumerate}
\item{$\F$ has a topology determined by a countable family of semi-norms.}
\item{$A_1,\, A_2, \dots , A_N$ are esa and are 
$\tauf$-continuous on $\F$.}
\item{For any $(\lambda_1 ,\, \lambda_2 , \dots ,\lambda_N )$ in 
$\Lambda$ there exists a generalized eigenvector in $\F^\times$, 
$|F_\lambda\rangle =|\lambda_1 , \lambda_2 , \dots ,\lambda_N \rangle$ 
such that

\begin{enumerate}
\item{$A^\times_i |F_\lambda \rangle = 
\lambda_i |F_\lambda \rangle$ for 
almost (with respect to $\mu$) all $i = 1,2, 
\dots , N$.  $A^\times_i$ denotes the extension of $A_i$ to $\F^\times$, the 
dual space to$\F$.  If the $A_i$ have no singular spectrum, ``almost all''
can be replaced by ``all''.}
\item{For any pair of vectors $\phi , \psi \in\F$ and any well defined 
function $f$ of $N$ variables, one has}

\end{enumerate}}
\end{enumerate}

\begin{equation}
 \left(  \phi ,  \psi \right) = 
 	 \int_\Lambda 	\langle \phi  | F_\lambda  \rangle 
		\langle F_\lambda |\psi\rangle 	\; 
			d\mu (\lambda_1 ,\, \lambda_2 , \dots ,\lambda_N )
\label{eq:partunity}
\end{equation}
\begin{equation}
 \left(  \phi , f(A_1,A_2,\dots ,A_N ) \psi \right) = 
 	\int_\Lambda f(\lambda_1 ,\, \lambda_2 , \dots ,\lambda_N )  
		\langle \phi |F_\lambda \rangle       
			\langle F_\lambda | \psi \rangle 	\; 
			d\mu (\lambda_1 ,\, \lambda_2 , \dots ,\lambda_N )
\label{eq:rspectres}
\end{equation}
The positive measure is unique up to equivalence of the null set.  In
general, the space $\F$ of the RHS may not be unique~\cite{note12}.


\begin{thebibliography}{00}
\bibitem{note3} 
{We have introduced the notion of correlation into a system in 
which long range interactions may figure.  Our methods are based on
quasi-invariant measures.  The Gibbs ensembles based 
on invariant measures are also compatible with long range
interactions.}
\bibitem{note10} 
{The extension of the $e^{-iHt}$ time evolution group on $\HH$ to 
transformations of the same $U(1)$ form on $\F$ and $\F^\times$ is not
a trivial matter, and more is involved than the generalization of the
form of formal solutions to the Schr{\"o}dinger equation.  It is a
mathematically and physically reasonable step to do so (discussed in
\cite{bgm} and~\cite{bmlg}), but one of the lessons of the present
work is that even 
a time evolution operator of apparently ``non-unitary'' form on
$\F\subset\F^\times$ could be ergodic and induce a unitary time
evolution operator on $\HH$ (in effect inducing the Schr{\"o}dinger
description of time evolution for pure states in $\HH$).  This will be
addressed in the third installment of this series~\cite{III}.}
\bibitem{note11} 
{We will not here address the issue of whether the bottom line of
  this idiosynchratic diagram can 
	simultaneously serve as the bottom level of a conventional 
	Gadella diagram, in effect determining the complex spaces 
	above this bottom line.}
\bibitem{note12}
{This statement of the complex spectral theorem stems
  from~\cite{bmlg} and~\cite{bgm}.  The present treatment is
  distinguishable in a variety 
       of ways, and one of the consequences of those distinctions is
       that our work involves different left and right
       (quasi-)invariant measures.  Our partition of unity is of the
       form $\II = \Sigma \, \vert F_\lambda \rangle \langle
       {\tilde{F_\lambda}} \vert$, where $\vert F_\lambda \rangle$ and
       $\langle {\tilde{F_\lambda}} \vert$ have different time domains
       of definition.  This will be addressed in the second and third
       installments of this series~\cite{II,III}, and is the result of
       attention to boundary conditions.}
\end{thebibliography}
\end{document}